\documentstyle[12pt]{article}          

\title{%
  Renormalization of \\
  Heavy Quark Effective Field Theory: \\
  Quantum Action Principles \\
  and Equations of Motion}

\author{%
  Wolfgang Kilian%
    \thanks{e-mail: {\tt Wolfgang.Kilian@Physik.TH-Darmstadt.de}}\\
  Thorsten Ohl%
    \thanks{e-mail: {\tt Thorsten.Ohl@Physik.TH-Darmstadt.de}}\\
  \hfill\\
    Technische Hochschule Darmstadt\\
    Schlo\ss{}gartenstr. 9\\
    D-64289 Darmstadt\\
    Germany\\
  \hfill\\
  IKDA 94/6\\
  hep-ph/9404305
  }

\date{April 1994}

\begin{document}
\maketitle

\begin{abstract}
  We discuss the quantum action principles and equations of motion for
  Heavy Quark Effective Field Theory. We prove the so-called
  equivalence theorem for HQEFT which states that the physical
  predictions of HQEFT are independent from the choice of
  interpolating fields.  En passant we point out that HQEFT is in fact
  more subtle than the quantum mechanical Foldy-Wouthuysen
  transformation.
\end{abstract}

\section{Introduction}
\label{sec:intro}

Heavy Quark Effective Field Theory (HQEFT) has arguably been the
fastest growing sector of elementary particle theory in the four years
since its introduction~\cite{HQET:Pioneers}.  This can be ascribed to
two features: on one hand the spin and flavor symmetries are
phenomenologically successful~\cite{HQET:Reviews} and, on the other
hand, corrections to the symmetry limit~$m_Q\to\infty$ are well
defined and therefore calculable.
Given this state of affairs, it seems appropriate to investigate the
theoretical formalism of HQEFT in all orders of the combined~$1/m_Q$,
$\alpha_S$ expansion.

It is only a slight exaggeration that Quantum Field Theory (QFT) has
no physical content and is just the most powerful tool known for the
calculation of~$S$-matrix elements~\cite{Weinberg:physica96a}.  The
fields and the off-shell values of their Green's functions are not
observable and therefore all QFTs which lead to identical~$S$-matrices
are physically equivalent.  We are therefore free to choose the
calculationally most convenient set of fields~\cite{Ge91}, similar to
the choice of appropriate coordinates in classical physics.  This
observation has borne the extremely powerful concept of Effective
Field Theory~(EFT)~\cite{EFT} which underlies almost all
phenomenologically successful applications of QFT.

It is however far from obvious how the general
theorems~\cite{Haag/Ruelle/Borchers} guaranteeing the independence of
the~$S$-matrix from the choice of interpolating fields translate to
the intricate formalism of renormalized perturbation
theory~\cite{CCWZ69}.
Nevertheless, the so-called equivalence theorem, which formalizes
this independence, has been proved rigorously~\cite{BL76},
using the renormalized quantum action principles~\cite{QAP,BM77}.  We
shall extend this proof to HQEFT below.

In order to avoid any misconceptions, we should stress that the
(colored) heavy quark fields~$h_v$, $\bar h_v$ of HQEFT are {\em not\/}
interpolating fields for~$S$-matrix elements in non-perturbative QCD.
This r\^ole is played
by the meson and baryon fields.  However, it seems instructive to
define the $S$-matrix for free quarks as an intermediate step
in order to show how the equivalence theorem works in a perturbative
calculation.

This note is organized as follows: to fix our notation and to
introduce the technical subtleties of higher order calculations, we
briefly describe HQEFT in section~\ref{sec:HQEFT}.  The quantum action
principles are introduced in section~\ref{sec:QAP}.  We sketch the
proof of the equivalence theorems for Green's functions and~$S$-matrix
elements in sections~\ref{sec:ET4GF} and~\ref{sec:ET4SM},
respectively. As an application, we discuss the reparametrization
invariance of HQEFT in section~\ref{sec:reparam} and clarify the
r\^ole of the classical equations of motion in section~\ref{sec:eom}.
Finally, we present our conclusions in section~\ref{sec:concl} and
describe why HQEFT is more subtle than the quantum mechanical
Foldy-Wouthuysen transformation.

\section{Heavy Quark Effective Field Theory}
\label{sec:HQEFT}

Within the framework of perturbation theory, the heavy mass expansion
is constructed by performing the following steps on the Green's
functions, i.e., Feynman diagrams of QCD:
Heavy quark loops are integrated out and replaced by their expansion
around vanishing momenta entering the loop.
The momenta of the remaining heavy quark propagators (those
connected to external heavy quark lines) are replaced by the residual
momenta~$k^\mu = p^\mu - mv^\mu$, $v^2=1$, $v_0>0$, with~$m$ the pole
mass of the heavy quark.
The antiparticle pole is eliminated by expanding the denominators
around~$k=0$:
\begin{equation}
	  \frac{1}{v\cdot k + k^2/2m+i\epsilon}
	  = \frac{1}{v\cdot k+i\epsilon}
	  \sum_{n=0}^\infty\left(
            \frac{-k^2/2m}{v\cdot k+i\epsilon}\right)^n.
\end{equation}

Although this procedure is legitimate for tree level amplitudes and for
the regularized loop diagrams, the different high energy behaviour of
the propagators
\begin{equation}
  G(k) =  i\frac{m{v\hskip -0.5em/} + {k\hskip-0.5em/} + m}{(mv +
     k)^2-m^2+i\epsilon}
\end{equation}
vs.
\begin{equation}\label{eq:S+}
  S_v(k) = \frac{iP_v^+}{v\cdot k + i\epsilon}
\end{equation}
in general destroys the equality of the full and effective theory
amplitudes after renormalization.  One has to correct for this
by introducing matching contributions into the effective theory.  This
amounts to calculating the differences of the full and effective
theory amplitudes
for any amputated one-particle irreducible (1PI) diagram of the full
theory
and absorb them into a set of local operator insertions in the effective
theory.

The matching works here order by order in the loop expansion because
the subtraction of the expansion about zero residual momentum
$k_i^\mu=0$ improves the infrared~(IR) behavior of the integrands,
exactly in such a way as to render the corrections IR convergent
\cite{KM94}.  Note that it is essential for $v$ to be timelike for
this to happen.

A tree level, the above procedure may be carried out in the functional
integral of QCD by integrating over the lower components of the heavy
quark field with respect to $v$ and thus constructing an effective
field theory (HQEFT).  The result \cite{MRR92}
\begin{eqnarray}\label{L-tree}
  {\cal L}^{(0)}_v &=&
  \bar h_v i(v\cdot D)h_v
  + (\bar h_viD_v^\perp + \bar R_v)
    \frac{1}{2m + i(v\cdot D)-i\epsilon}
    (iD_v^\perp h_v + R_v)  \nonumber\\
  && +\; \bar\rho_v h_v + \bar h_v\rho_v,
\end{eqnarray}
containing the transverse derivative
\begin{equation}
  D_v^\perp = {D\hskip-0.6em/} - {v\hskip -0.5em/}(v\cdot D).
\end{equation}
is expressed in terms of the upper component field
\begin{equation}
  h_v(x) = e^{imv\cdot x} P_v^+ Q(x)
	\quad\quad\left[P_v^\pm=(1\pm{v\hskip-0.5em/})/2\right]
\end{equation}
and the sources of upper and lower components
\begin{eqnarray}
  \rho_v(x) &=& P_v^+ e^{imv\cdot x}\eta(x), \\
  R_v(x)    &=& P_v^- e^{imv\cdot x}\eta(x).
\end{eqnarray}
The latter are not needed for deriving $S$-matrix elements, but one
might find it convenient to retain them in order to
easily obtain the effective theory analogues of operator insertions.

Matching contributions enter in the same way as the tree-level
quark-gluon vertex, so they can by included into (\ref{L-tree}) by
introducing a generalized covariant derivative \cite{KM94}
\begin{equation}\label{D-gen}
  i{{\cal D}\hskip-0.6em/} = i{D\hskip-0.6em/} + O(\alpha_s)
\end{equation}
with projections
\begin{eqnarray}\label{D-gen-p}
  i{\cal D}_v^+ &=& P_v^+i{{\cal D}\hskip-0.6em/} P_v^+ = (iv\cdot
     D)P_v^+ + O(\alpha_s),\nonumber\\
  i{\cal D}_v^\perp &=& P_v^+i{{\cal D}\hskip-0.6em/} P_v^- +
     P_v^-i{{\cal D}\hskip-0.6em/} P_v^+
	= iD_v^\perp + O(\alpha_s)\nonumber,\\
  i{\cal D}_v^- &=& P_v^-i{{\cal D}\hskip-0.6em/} P_v^- = -(iv\cdot
     D)P_v^- + O(\alpha_s).
\end{eqnarray}
The effective Lagrangian which generalizes (\ref{L-tree}) is
\begin{eqnarray}\label{L-loop}
  {\cal L}_v &=&
  \bar h_v i{\cal D}_v^+ h_v
  + (\bar h_vi{\cal D}_v^\perp + \bar R_v)
    \frac{1}{2m-i{\cal D}_v^- }
    (i{\cal D}_v^\perp h_v + R_v) +{\cal C}_0 \nonumber\\
  && +\;\bar\rho_v h_v + \bar h_v\rho_v,
\end{eqnarray}
where~${\cal C}_0$ summarizes operators consisting only of light
fields which are introduced by integrating out heavy quark loops in
the full theory.  This effective Lagrangian is valid to arbitrary (but
finite) order in the $1/m$ and loop expansions.

One can easily reformulate HQEFT with an arbitrary residual mass
$\delta m$ of the order $\Lambda_{\rm QCD} \ll m$~\cite{FNL92} by the
change $D \to D - \delta m\,v$ and the induced changes in ${\cal D}$.
For generality, we include this residual mass term in the heavy quark
propagator.

\section{Quantum Action Principles}
\label{sec:QAP}

It is useful to establish some notation for a concise presentation of
the Quantum Action Principles (QAP).  For convenience, we shall
denote all fields, heavy and light, generically by a scalar
field~$\phi$.  The generating functional of the renormalized Green's
functions is given by
\begin{equation}
  Z[j]_{\cal L} = \sum_{n=0}^\infty \frac{i^n}{n!}
	\int dx_1\ldots dx_n j(x_1) \ldots j(x_n)
        \left\langle 0 \left\vert {\rm T}\left[
           \phi(x_1)\ldots\phi(x_n) \right] \right\vert 0
        \right\rangle_{{\cal L}(\phi)}
\end{equation}
and we use the following shorthand for local operator insertions:
\begin{equation}
  {\cal O}(z) \Downarrow Z[j]_{\cal L}
     = \sum_{n=0}^\infty \frac{i^n}{n!}
	\int dx_1\ldots j(x_1) \ldots
        \left\langle 0 \left\vert {\rm T}\left[ {\cal
          O}(z)\phi(x_1)\ldots \right] \right\vert 0
          \right\rangle_{{\cal L}(\phi)}.
\end{equation}
By including appropriate source terms into the lagrangian, we can
express the generating functionals formally as functional integrals
\begin{eqnarray}
\label{eq:Z-pi}
  Z[j]_{\cal L}
    & = & \int{\cal D}\phi\; e^{i\int{\cal L}(\phi;j)} \\
\label{eq:OZ-pi}
  {\cal O}(z) \Downarrow Z[j]_{\cal L}
    & = & \int{\cal D}\phi\; {\cal O}(z)\,
        e^{i\int{\cal L}(\phi;j)},
\end{eqnarray}
but these expressions have to be renormalized properly.

Using formal manipulations of the functional integrals~(\ref{eq:Z-pi})
and~(\ref{eq:OZ-pi}) we can derive the three quantum action principles
for the generating functionals immediately:
\begin{enumerate}
  \item invariance under variations of the quantized fields~$\phi$,
    i.e.~field equations:
    \begin{equation}
    \label{eq:QAP1}
      \delta {\cal L}(z) \Downarrow Z[j]_{\cal L} = 0, \quad
      \delta {\cal L}(\phi)
         = \frac{\delta {\cal L}(\phi)}{\delta \phi} \delta \phi
    \end{equation}
    and
    \begin{equation}
      \delta \phi(x) = \left(P(\phi)\right)(x) \,\delta \epsilon
    \end{equation}
    with $P(\phi)$ an arbitrary local polynomial in $\phi$ and its
    derivatives.
  \item change under variation of external fields:
    \begin{equation}
    \label{eq:QAP2}
      \frac{\delta {\cal L}}{\delta \chi}(x) \Downarrow
         Z[j;\chi]_{\cal L}
         = -i \frac{\delta Z[j;\chi]_{\cal L}}{\delta \chi}(x)
    \end{equation}
  \item change under variation of parameters in the lagrangian
    \begin{equation}
    \label{eq:QAP3}
      \frac{\partial {\cal L}}{\partial \eta} \Downarrow Z[j]_{\cal
         L(\eta)}
         = -i \frac{\partial Z[j]_{\cal L(\eta)}}{\partial \eta}.
    \end{equation}
\end{enumerate}
After renormalization the three QAPs~(\ref{eq:QAP1}), (\ref{eq:QAP3}),
and~(\ref{eq:QAP3}) are valid without changes if the dimensional
renormalization scheme is used~\cite{HV72,BM77}.
In other renormalization schemes (e.g.~BPHZ~\cite{BPHZ})
there are (calculable) normal product corrections~\cite{QAP}.

Since dimensional regularization is defined with the help of
the Schwinger parameter representation, which is also convenient
for practical calculations, we define the parameter representation
for diagrams containing heavy particles:

Any internal light line $\ell$ is assigned a parameter
$\alpha_\ell$ and an auxiliary vector $u_\ell$.  The
propagator is represented by
\begin{equation}
  \frac{iD(k_\ell)}{k_\ell^2-m_\ell^2+i\epsilon}
  = \left.\int_0^\infty d\alpha_\ell\,
    D(-i\partial/\partial u_\ell)
    \exp i(\alpha_\ell k_\ell^2 + u_\ell\cdot k_\ell
	- \alpha_\ell [m_\ell^2-i\epsilon])\right|_{u_\ell=0}.
\end{equation}
where $D(k_\ell)$ may contain algebraic objects like
$\gamma$ matrices, Lorentz tensors etc.
Any internal heavy line $h$ is assigned a parameter
$\beta_h$ and an auxiliary vector $u_h$.  The
propagator is represented by
\begin{equation}\label{eq:beta}
  \frac{iP_v^+}{v_h\cdot k_h - \delta m_h +i\epsilon}
  = P_v^+\left.\int_0^\infty d\beta_h\,
    \exp i([\beta_h v_h + u_h]\cdot k_h - \beta_h[\delta m_h -
      i\epsilon])\right|_{u_h=0}.
\end{equation}
Any vertex factor $V(p_i,k_i)$ with $p_i$, $k_i$ being the
momenta of the attached external and internal lines, respectively,
is translated into $V(p_i,-i\partial/\partial u_i)$.
Finally, the Gaussian momentum integrations are formally carried out.

The requirement that no heavy particle loops are included ensures
that all momentum integrations are indeed Gaussian.  In the following
we consider only the case of one heavy quark with velocity $v$.
The generalization to the multiparticle sector is straightforward
if the velocities of the heavy particles do not coincide.

Here we shall {\em not\/} present a proof of theorem I of~\cite{BM77},
i.e.~the consistency of dimensional renormalization for HQEFT in the
parametric representation.  We do not expect a failure of a
translation of the proof in~\cite{BM77}, and we hope to come back to
the purely technical details of such a proof in a later publication.

In order to establish the first QAP (\ref{eq:QAP1}) in HQEFT, we
have to specify the separation of the Lagrangian ${\cal L}$ into
the free (${\cal L}_0$) and interaction (${\cal L}_{\rm int}$) parts.
We take ${\cal L}_0$ to be equal to
\begin{equation}
  {\cal L}_0 = \bar h_v (iv\cdot\partial - \delta m) h_v
\end{equation}
corresponding to the propagator (\ref{eq:S+}).  The QAP (\ref{eq:QAP1})
then says
\begin{equation}\label{eq:QAP1-eff}
  \left(\bar h_v P(\phi)\, (iv\cdot\partial - \delta m)h_v\right)(z)
	\Downarrow Z[j]_{\cal L}
  = -\left(\bar h_v P(\phi)\,\frac{\delta{\cal L}_{\rm
       int}}{\delta\bar h_v}\right)(z)
	\Downarrow Z[j]_{\cal L}
\end{equation}
where now $\phi$ denotes the light fields of the theory only.
In order to remain in the one heavy particle sector, we insist that
the field variations be linear in the heavy quark field.

In diagrammatic language, the content of (\ref{eq:QAP1-eff}) is that
attaching the factor $(iv\cdot\partial - \delta m)$ to a vertex is
equivalent to contracting the neighbouring heavy quark line to a
point.  In the parametric representation, we have to show that
\begin{equation}
  \left(v\cdot\frac{\partial}{\partial u_h} - i\,\delta m\right)
	I(p,u,\alpha,\beta) =
  \frac{\partial}{\partial\beta_h} I(p,u,\alpha,\beta)
\end{equation}
which is true since apart from a global factor $\exp -i\sum_h
\beta_h\,\delta m$, the dimensionally regularized integrand $I$
depends on $u_h$ and $\beta_h$ only in the combination $\beta_h v +
u_h$.  The desired equality follows from the fact that
\begin{equation}
  \int_0^\infty d\beta_h\frac{\partial}{\partial\beta_h}I
	=  -I(\beta_h=0),
\end{equation}
since $I$ falls off exponentially at the upper limit.

The proof that the QAP remains valid after renormalization proceeds
exactly as in \cite{BM77}.  The only potential obstruction are diagrams
such as depicted in Fig.~1a, where the subdiagram
enclosed by dashed lines has no equivalent if the center line of the
diagram is contracted to a point (cf.~also Fig.~1 of~\cite{BM77}).
The corresponding counterterm diagram would contribute to the left-hand
side of (\ref{eq:QAP1-eff}), but not to the right-hand side.  However,
this counterterm diagram (Fig.~1b) necessarily
contains a heavy quark loop and thus vanishes identically.

The proof of the second QAP~(\ref{eq:QAP2}) is simple combinatorics
since ${\cal L}_0$ is independent of the external fields.  The same
applies to the third QAP (\ref{eq:QAP3}, \ref{eq:QAP3-eff}) if ${\cal
L}_0$ is independent of the parameter $\eta$.  Therefore we shall
restrict ourselves to the case where the varied parameter is the
velocity of the heavy quark
\begin{equation}
  v \to v+ \eta\,\delta v\quad\mbox{with}\quad v\cdot\delta v=0;
\end{equation}
the case $\eta=\delta m$ can be treated similarly.

We have to prove:
\begin{equation}\label{eq:QAP3-eff}
  \frac{\partial}{\partial\eta} Z[j]_{\cal L}
  =  - \int dz (\bar h_v \delta v\cdot\partial h_v)(z)
	\Downarrow Z[j]_{\cal L}
	+ i \frac{\partial{\cal L}_{\rm int}}{\partial \eta}
	\Downarrow Z[j]_{\cal L}.
\end{equation}
The second term on the right-hand side is cancelled exactly by the terms
on the left-hand-side where the derivative acts on the vertices.
Furthermore, acting with $\partial/\partial\eta$
on the propagator part of the amplitude results in terms of the form
\begin{eqnarray}
  \lefteqn{\frac{\partial}{\partial\eta}
  \left.\int_0^\infty d\beta\,
  I(u+\beta[v+\eta\,\delta v],\beta\,\delta m,\ldots)\right|_{u=0}}
  \nonumber\\
  &=& \left.\int_0^\infty d\beta\,\beta\,
	\delta v\cdot\frac{\partial}{\partial u}
  I(u+\beta[v+\eta\,\delta v],\beta\,\delta m,\ldots)\right|_{u=0} \\
  &=& \left.\int_0^\infty d\beta\,d\beta'\delta v\cdot
  \frac{\partial}{\partial u'}
  I(u+\beta[v+\eta\,\delta v]+u'+\beta'[v+\eta\,\delta
     v],(\beta+\beta')\delta m,\ldots)
	\right|_{\stackrel{\scriptstyle u=0}{u'=0}}.\nonumber
\end{eqnarray}
The last expression corresponds to a Feynman diagram where two heavy
lines are joined by the quadratic vertex~$-\bar h_v \delta
v\cdot\partial h_v$,
which is equivalently obtained from the first term of the right-hand
side of (\ref{eq:QAP3-eff}).  Simple combinatorics completes the proof
for the regularized amplitudes, and the validity of
(\ref{eq:QAP3-eff}) after renormalization is proved as described
in \cite{BM77}.

\section{The Equivalence Theorem for Green's Functions}
\label{sec:ET4GF}

The equivalence theorem asserts the following
relations among Green's functions
\begin{equation}
\label{eq:gf}
  \left\langle 0 \left\vert {\rm T}\left[ \phi(x_1)\ldots\phi(x_n)
    \right] \right\vert 0 \right\rangle_{{\cal L}(\phi)}
  = \left\langle 0 \left\vert {\rm T}\left[ \phi'(x_1)\ldots\phi'(x_n)
    \right] \right\vert 0 \right\rangle_{{\cal L'}(\phi)={\cal
      L}(\phi')}
\end{equation}
of fields which are related by a reparametrization
\begin{equation}
  \phi(x) \to \phi'(x) = \phi(x) + \eta F(\phi(x))
\end{equation}
in all orders of a simultaneous expansion in the numbers of loops and
powers of~$\eta$.

We concentrate on field reparametrizations which are linear in
the heavy field $\bar h_v$, but may be nonlinear in the other fields of
the theory:
\begin{equation}\label{eq:hv-prime}
  \bar h_v(x) \to \bar h_v'(x) = \bar h_v(x)\left( 1 + \eta
     F(\phi(x),\partial)\right).
\end{equation}
The structure of the heavy quark Lagrangian allows to partially
integrate
so that the derivatives in $F$ act on~$h_v$ and the light fields only.
This simplifies
the notation in the following arguments.  Of course, the proof can
be applied to $h_v$ also.  Simultaneous changes in $h_v$ and $\bar h_v$
may be splitted into subsequent reparametrizations of $h_v$ and $\bar
h_v$.

We want to show
\begin{equation}\label{eq:eqt-gf}
  \left\langle 0 \left\vert {\rm T}\left[ \bar h_v(x)\ldots \right]
    \right\vert 0 \right\rangle_{{\cal L}(\bar h_v,h_v)}
  = \left\langle 0 \left\vert {\rm T}\left[ \bar h_v'(x)\ldots \right]
      \right\vert 0 \right\rangle_{{\cal L}'(\bar h_v, h_v) = {\cal
      L}(\bar h_v',h_v)}.
\end{equation}
where it is understood that the Green's functions are evaluated to
some finite
order $\eta^n$.  Following Lam \cite{BL76}, we define
\begin{equation}
  Z(\eta) = Z[j]_{{\cal L}(\bar h_v + \eta \bar
        h_vF(\phi,\partial),h_v)}.
\end{equation}
The first QAP (\ref{eq:QAP1}) tells us that
\begin{equation}
  \bar h_v P(\phi)\frac{\delta{\cal L}'}{\delta\bar h_v}\Downarrow
    Z(\eta) = 0.
\end{equation}
for any local polynomial~$P(\phi)$ in the light fields.  From
(\ref{eq:hv-prime}) this is equivalent to
\begin{equation}
  \bar h_v P(\phi)\frac{\delta{\cal L}'}{\delta\bar h_v'}
	\Downarrow Z(\eta)
  = -\eta\, \bar h_v P(\phi)\,F(\phi)\frac{\delta{\cal L}'}{\delta\bar
     h_v'} \Downarrow Z(\eta).
\end{equation}
After iterating this equation $n$~times by replacing~$P(\phi)$
by~$P(\phi)\,F(\phi)^k$ for~$k=1,2,\ldots n$ on the left-hand side,
we arrive at
\begin{equation}\label{eq:eta-n}
  \bar h_v P(\phi)\frac{\delta{\cal L}'}{\delta\bar h_v'}
	\Downarrow Z(\eta)
  = (-\eta)^{n+1} \bar h_v P(\phi)\,F(\phi)^{n+1}\frac{\delta{\cal
      L}'}{\delta\bar h_v'} \Downarrow Z(\eta).
\end{equation}
Since we are interested in the variation of~$Z(\eta)$ with~$\eta$, we
use the third QAP~(\ref{eq:QAP3}) to get
\begin{equation}
  \frac{d}{d\eta} Z(\eta)
  =  i\frac{\partial{\cal L}'}{\partial\eta}\Downarrow Z(\eta)
  =  i\bar h_v F(\phi) \frac{\delta {\cal L}'}{\delta\bar
     h_v'}\Downarrow Z(\eta).
\end{equation}
Applying~(\ref{eq:eta-n}) with~$P=F$, we see that the generating
functional~$Z(\eta)$ is independent of~$\eta$ in any finite order~$n$:
\begin{equation}
  \frac{d}{d\eta} Z(\eta)
  =  O(\eta^{n+1}).
\end{equation}
Taking derivatives with respect to the sources completes the proof of
the equivalence theorem (\ref{eq:eqt-gf}) for renormalized Green's
functions in the effective theory.  Operator insertions as those
responsible for weak decays may also be considered by coupling them to
some source.

We emphasize that this proof holds to finite order in $\eta$ and in the
dimensional renormalization scheme (MS or $\overline{\rm MS}$) only.
In other
renormalization schemes (e.g.\ BPHZ, lattice) there are corrections to
(\ref{eq:eqt-gf}).  The same is true for resummed variations which
modify the propagator.

\section{The Equivalence Theorem for the $S$-Matrix}
\label{sec:ET4SM}

After discussing the proper definition of~$S$-matrix elements in
HQEFT, we shall derive the equivalence theorem for these~$S$-matrix
elements.  In perturbation theory (or for that matter in any
non-confining theory like QED), there are asymptotic states
corresponding to single heavy particles.  Translating the familiar
LSZ reduction formula for a massive fermion in momentum space
\begin{equation}\label{LSZ-ft}
  S(p,\ldots) = \frac{-i}{\sqrt{Z_Q}}\bar u(p/m)\,({p\hskip-0.5em/} -
                 m)\cdots
	\left.\int dx\,e^{ip\cdot x}{\left\langle 0 \left\vert {\rm
            T}\left[ Q(x)\cdots \right] \right\vert 0
            \right\rangle}\right|_{p^2=m^2}
\end{equation}
to the effective theory, and setting $\delta m=0$, we use
\begin{equation}\label{ubar}
  \bar u(v+k/m) = \bar u(v)\frac{{v\hskip -0.5em/} + 1 +
                  {k\hskip-0.5em/}/m}
	{2\sqrt{ 1 + (v\cdot k)/2m}}
\end{equation}
to arrive at
\begin{eqnarray}\label{LSZ-eft}
  \lefteqn {S(k,\ldots)} \\
  & = & \frac{-i}{\sqrt{Z_Q \tilde Z(k)}}\bar u(v)
	\left(v\cdot k + \frac{k^2}{2m}\right)
	\left.\int dx\,e^{ik\cdot x}{\left\langle 0 \left\vert {\rm
        T}\left[ h_v(x)\cdots \right] \right\vert 0 \right\rangle}
	\right|_{v\cdot k = -k^2/2m}.  \nonumber
\end{eqnarray}
Here the kinematical factor
\begin{equation}\label{eq:Z(k)}
  \tilde Z(k) = 1 + \frac{v\cdot k}{2m}
\end{equation}
follows from the matching condition~(\ref{ubar}) for the heavy
particle wave functions.  Alternatively, we can derive it from the
residue of the heavy particle two-point function~${\left\langle 0
\left\vert {\rm T}\left[ h_v(x)\bar h_v(y) \right] \right\vert 0
\right\rangle}$ at the one heavy particle pole.  The additional
term~$k^2/2m$ in (\ref{LSZ-eft}) cancels the multiple poles in the
heavy-quark Green's functions that arise from insertions of the $1/m$
terms in the Lagrangian into external lines.

Having defined the $S$-matrix for heavy quark states, the proof that
the $S$-matrix is unchanged by reparametrizations, i.e., that the
$S$-matrix elements calculated from the original HQEFT are the same as
those calculated with the new Lagrangian obtained by the change of
variables (\ref{eq:hv-prime}), proceeds exactly as in ordinary quantum
field theory~\cite{BL76}.  The additional terms in the Green's
functions on the right-hand side of (\ref{eq:eqt-gf}) contribute in
such a way that the residue of the pole at $v\cdot k + k^2/2m=0$
(which appears as a series of multiple poles in the effective theory)
is multiplied by some function of $k$ identical for all Green's
functions of the heavy quark.  This function only modifies the factor
$\tilde Z(k)$ in (\ref{LSZ-eft}) and therefore does not contribute to
$S$-matrix elements.  Any contribution which has no pole at $v\cdot k
+ k^2/2m=0$ vanishes after applying (\ref{LSZ-eft}).

In the real world heavy quarks come in bound states, and the $S$-matrix
is defined in terms of those.  Denoting a heavy (scalar) bound state
with mass $M$ generically by $B$, a typical matrix element is given by
\begin{eqnarray}\label{LSZ-B}
  \left\langle B \right\vert {\cal O} \left\vert X \right\rangle
  &=& \frac{1}{\sqrt{Z_B}} \left\langle 0 \right\vert J_B \left\vert B
     \right\rangle\left\langle B \right\vert {\cal O}\left\vert X
     \right\rangle \\
  &=& \frac{-i}{\sqrt{Z_B}}\left(P^2-M^2\right)\int dx\,e^{iP\cdot
        x}\left.
        \left\langle 0 \left\vert {\rm T}\left[J_B(x)\,{\cal
        O}(0)\right]
           \right\vert X \right\rangle
	   \right|_{P^2=M^2}
	\nonumber
\end{eqnarray}
where
\begin{equation}
  Z_B = \left(P^2-M^2\right)\int dx\,e^{iP\cdot x}\left.
	{\left\langle 0 \left\vert {\rm T}\left[ J_B(x)\,J_B(0)
        \right] \right\vert 0 \right\rangle}\right|_{P^2=M^2}
\end{equation}
and $J_B$ is any operator such that~$\left\langle 0 \left\vert J_B(x)
\right\vert B \right\rangle$ does not vanish.  In the
effective theory we have
\begin{eqnarray}\label{LSZ-B-eff}
  \left\langle B \right\vert{\cal O}\left\vert X \right\rangle
  &=&  \frac{-i}{\sqrt{Z_B^{\rm eff}}}\left(v\cdot k + \frac{k^2}{2m} -
      \frac{M^2-m^2}{2m}\right) \\
  && \times \int dx\,e^{ik\cdot x}\left.
	\left\langle 0 \left\vert {\rm T}\left[J_B^{\rm eff}(x)\,{\cal
           O}(0)\right] \right\vert X \right\rangle
	\right|_{v\cdot k + \frac{k^2}{2m} = \frac{M^2-m^2}{2m}},
	\nonumber
\end{eqnarray}
where $J_B^{\rm eff}$ is some interpolating field of the effective
theory, and the definition of $Z_B$ is modified analogously
\begin{eqnarray}
  Z_B^{\rm eff} & = & \left(v\cdot k + \frac{k^2}{2m} -
     \frac{M^2-m^2}{2m}\right) \nonumber\\
     & & \times \int dx\,e^{ik\cdot x}\left.
         \left\langle 0 \left\vert {\rm T}\left[ J_B^{\rm
          eff}(x)\,J_B^{\rm eff}(0) \right] \right\vert 0 \right\rangle
	\right|_{v\cdot k + \frac{k^2}{2m} = \frac{M^2-m^2}{2m}}.
\end{eqnarray}
The difference of hadron and quark masses has to be treated in the same
perturbative expansion
\begin{equation}
  M = m + \bar\Lambda + \frac{\lambda}{2m} + \ldots
\end{equation}
so that multiple poles in the matrix element due to the mass shift
are cancelled separately in any order of the $1/m$ expansion.  Of
course, one could rearrange the series to be an expansion in $1/M$
or any other physically sensible mass by suitable choice of $m$.

Using similar arguments as for the free quark case, it becomes obvious
that any reparametrization of the heavy quark field that changes the
form of the interpolating field $J_B$ has no effect on the ratio
(\ref{LSZ-B-eff}).  Since the choice of $J_B$ has been arbitrary from
the beginning, one might as well keep the definition of $J_B$ fixed
while calculating a matrix element using the reparametrized
Lagrangian.  The equivalence theorem ensures that masses and other
properties of bound states remain unchanged.  However, since
nonperturbative QCD cannot be formulated using the MS or
$\overline{\rm MS}$
scheme, in order to calculate a particular matrix element one first
has to match renormalization schemes (e.g.\ lattice versus
$\overline{\rm MS}$).
In a general scheme the equivalence theorem does not hold without
corrections, so one will in general obtain different matching
contributions in different versions of the effective theory.

\section{Reparametrization Invariance}
\label{sec:reparam}

As an application of the QAPs and the equivalence theorem, we discuss
the implication of the Luke-Manohar reparametrization invariance
\cite{LM92} on the effective Lagrangian.  For simplicity, we keep
$\delta m$ equal to zero in this section.

An infinitesimal change in the velocity
$v$ used in the derivation of (\ref{L-loop})
\begin{equation}
  v \to v+ \delta v \quad \mbox{with}\quad v\cdot\delta v = 0
\end{equation}
causes a corresponding change in the field $h_v$
\begin{equation}
  h_{v+\delta v} = e^{im(\delta v\cdot x)}
	P_{v+\delta v}^+ (h_v + H_v).
\end{equation}
Since the $H_v$ field is integrated out, this implies
\begin{equation}
  \delta h_v = \left(
	im(\delta v\cdot x) + \frac{\delta{v\hskip -0.5em/}}{2}
	+ \frac{\delta{v\hskip -0.5em/}}{2}
	\frac{1}{2m-i{\cal D}_v^-}i{\cal D}_v^\perp\right)h_v.
\end{equation}
Inserting this variation into the effective Lagrangian (\ref{L-loop}),
a straightforward calculation shows that ${\cal L}_v$ is
reparametrization invariant
\begin{equation}
  \delta{\cal L}_v = 0
\end{equation}
if and only if
\begin{equation}
  \delta(i{\cal D}_v + m{v\hskip -0.5em/})
  = - [i\tilde{\cal D}_v + m{v\hskip -0.5em/}, im(\delta v\cdot x)].
\end{equation}
Assuming gauge invariance, this condition is satisfied if and only if
the
generalized covariant derivative ${\cal D}_v$ depends on the ordinary
covariant derivative $D$ and the velocity $v$ in the combination
\begin{equation}
  i{\cal D}_v + m{v\hskip -0.5em/} = f(v+iD/m)
\end{equation}
This is exactly what one would expect since $mv+k$ is the full momentum
of the heavy quark, the quantity that enters into the matching
calculation. In particular, since
\begin{equation}
  i{D\hskip-0.6em/} = m({v\hskip -0.5em/} + i{D\hskip-0.6em/}/m) -
                      m{v\hskip -0.5em/},
\end{equation}
reparametrization invariance is valid on tree level, as it has been
shown already in~\cite{Ch93}.

The QAPs imply that the relations following from this type of
re\-pa\-ra\-met\-ri\-za\-tion invariance remain valid after
renormalization, if
the division of the effective Lagrangian into renormalized part
and counterterms is carried out using the MS or $\overline{\rm MS}$
renormalization scheme.  In other schemes reparametrization
invariance may be broken.

\section{Equations of motion}
\label{sec:eom}

The physical content of the first QAP (\ref{eq:QAP1}, \ref{eq:QAP1-eff})
is that naive application of the classical equations of motion
\begin{equation}
  (iv\cdot D - \delta m)h_v =
	-\frac{\delta}{\delta\bar h_v}{\cal L}^{(1)}
\end{equation}
is justified in the $\overline{\rm MS}$ renormalization scheme,
where~${\cal
L}^{(1)}$ denotes the terms of order~$1/m$ and higher in the effective
Lagrangian.  However, as it stands the QAP applies only to single
operator insertions such as weak currents.  If multiple operator
insertions~${\cal O}_i$ are considered, the easiest way to obtain the
correct terms is to couple sources~$j_i$ to them and include them in
the effective Lagrangian.  In higher orders of the loop expansion, 1PI
diagrams containing multiple insertions generate counterterms
nonlinear in the sources, so that the effective Lagrangian is some
polynomial in~$j_i$.  Employing the equivalence theorem, one can look
for a reparametrization of the heavy quark field that results in a
particular simplification.  If the Lagrangian contains an operator of
the form
\begin{equation}
  {\cal O} = \bar h_v F(\phi,iD)\,(iv\cdot D - \delta m) h_v,
\end{equation}
one may set
\begin{equation}\label{eq:rep-gen}
  h_v \to \left(1 - F(\phi,iD)\right) h_v.
\end{equation}
so that the variation of the lowest-order term ${\cal L}^{(0)}$
eliminates
${\cal O}$ in favor of a tower of new higher-dimensional operators.
This method may be used to eliminate all terms containing $(iv\cdot D-
\delta m)$ from the interaction Lagrangian.  However, the familiar
Luke-Manohar reparametrization invariance is obscured in this process.
Furthermore, the kinematical normalization factor $\tilde Z(k)$ in
(\ref{eq:Z(k)}) has to be recalculated, but this is only relevant for
free heavy quarks as external states.

In fact, any allowed reparametrization is of the form
(\ref{eq:rep-gen}) and thus equivalent to the application of the
equations of motion:  The additional operator introduced into the
Lagrangian is
\begin{equation}
  -\frac{\delta{\cal L}}{\delta h_v} F(\phi,iD) h_v.
\end{equation}
The equivalence theorem tells us that the vanishing of this operator
works for an arbitrary number of insertions, whereas the QAP applies
only to single insertions.

On the other hand, reparametrizations are as well possible in the full
theory, if the additional terms introduced in the QCD Lagrangian are
treated perturbatively.  Since the effective theory may be constructed
also from the reparametrized Lagrangian by calculating the
appropriate matching corrections, we have the following situation:
The equivalence theorem in the
full {\em and\/} effective theories, respectively, guarantees that the
matching contributions do not depend on in which theory
reparametrizations are carried out, if we use dimensional
renormalization in both cases.  In ordinary field theory one usually
considers
only field redefinitions that are relativistically invariant
\begin{equation}
  Q(x) \to Q(x) + F(\phi, {\partial\hskip-0.5em/})\,Q(x).
\end{equation}
Such redefinitions may be mirrored in the effective theory; we observe
that they preserve the Luke-Manohar reparametrization invariance of
the effective Lagrangian.  Other transformations correspond to
Lorentz non-invariant repara\-metri\-zations of the full theory.

\section{Conclusions}
\label{sec:concl}

In this letter we have established the quantum action principles
(QAPs) in HQEFT which justify formal manipulations of
the functional integral.  Using the QAPs, we have proved the
equivalence theorem for the effective theory and thus have shown under
what conditions field redefinitions in HQEFT are viable.  A particular
class consists of transformations corresponding to the Lorentz
invariance of QCD --- the Luke-Manohar reparametrization
invariance of HQEFT.  In practice, the equivalence theorem serves as a
guideline how the classical equations of motion manifest themselves
once radiative corrections and renormalization are taken into account.

{}From our discussion it should have become clear that --- unless one is
willing to handle normal product corrections --- it is necessary to
use a MS-like renormalization scheme, and to make sure that the free
Lagrangian (i.e., the propagator) is not affected.  In particular,
this applies to attempts to derive the heavy quark Lagrangian
employing the Foldy-Wouthuysen transformation familiar from
single-particle quantum mechanics \cite{KT91}.  By its very nature this
transformation modifies the quark propagator in the full theory and
thus violates the hypothesis of the equivalence theorem.  Since the
authors of~\cite{KT91} do not discuss how to calculate the necessary
normal product corrections, the argument given in \cite{KT91} is not
a complete derivation of HQEFT.  On the other hand, the Lagrangian
provided in~\cite{KT91} can be reduced to the tree-level
Lagrangian~(\ref{L-tree}) by a series of field redefinitions which are
compatible with the equivalence theorem of
the {\em effective\/} theory\footnote{These redefinitions change the
normalization of the heavy quark field.  The authors of~\cite{KT91}
have argued that the admissible reparametrizations have to preserve
the normalization of the fields.  While this is applicable to wave
functions in single-particle quantum mechanics, the particular
normalization of quantized fields cancels in the calculation
of~$S$-matrix elements in the full theory~(\ref{LSZ-ft}, \ref{LSZ-B})
as well as in he effective theory~(\ref{LSZ-eft}, \ref{LSZ-B-eff}).}.
Thus beyond tree-level the Foldy-Wouthuysen approach can
{\em only\/} be justified if it is amended by
some prescription to calculate matching corrections, which will then
yield the same results as the conventional approach~\cite{MRR92,KM94}.

To summarize, the form of the HQEFT Lagrangian is by no means unique,
but as long as a MS-like renormalization scheme is used, different
versions can be mapped onto each other by perturbative field
redefinitions.  However, in any renormalization scheme where the
QAPs hold in the weak sense only (i.e., normal product corrections
arise), one has to perform a new matching calculation if one wants to
carry out reparametrizations.  In particular, in a general scheme the
classical equations of motion cannot be naively applied, and
reparametrization invariance is lost beyond leading order in the
$\alpha_s$ and $1/m$ expansions.


\section*{List of Figures}
\begin{itemize}
  \item[1.] Potential obstruction in the proof of the QAP.
            The open square denotes the operator insertion on the
            left-hand side of~(23).

\end{itemize}

\end{document}